\documentclass[10pt,conference]{IEEEtran}
\IEEEoverridecommandlockouts
\usepackage{cite}
\usepackage{amsmath,amssymb,amsfonts}
\usepackage{algorithmic}
\usepackage{graphicx}
\usepackage{textcomp}
\usepackage{xcolor}
\usepackage{subfiles}
\usepackage{tabularx}
\usepackage{booktabs}
\usepackage[capitalise]{cleveref}
\usepackage{xspace}
\usepackage{url}
\usepackage{caption}
\usepackage{subcaption}
\usepackage{balance}
\usepackage{tabularx}
\def\BibTeX{{\rm B\kern-.05em{\sc i\kern-.025em b}\kern-.08em
    T\kern-.1667em\lower.7ex\hbox{E}\kern-.125emX}}

\newcommand{\toolname}{\emph{Code Critters}\xspace}

\begin{document}

\title{Code Critters: A Block-Based Testing Game}

\author{\IEEEauthorblockN{Philipp Straubinger}
\IEEEauthorblockA{\textit{University of Passau} \\
Passau, Germany}
\and
\IEEEauthorblockN{Laura Caspari}
\IEEEauthorblockA{\textit{University of Passau} \\
	Passau, Germany}
\and
\IEEEauthorblockN{Gordon Fraser}
\IEEEauthorblockA{\textit{University of Passau} \\
Passau, Germany}
}

\maketitle

\begin{abstract}
  Learning to program has become common in schools, higher education and individual learning.
  Although testing is an important aspect of programming, it is often neglected in education due to a perceived lack of time and knowledge, or simply because testing is considered less important or fun.
  To make testing more engaging, we therefore introduce \toolname, a Tower Defense game based on testing concepts: The aim of the game is to place magic mines along the route taken by small ``critters'' from their home to a tower, such that the mines distinguish between critters executing correct code from those executing buggy code. Code is shown and edited using a block-based language to make the game accessible for younger learners. The mines encode test inputs as well as test oracles, thus making testing an integral and fun component of the game.
\end{abstract}


\section{Introduction}

Programming has become a common aspect of education at schools~\cite{kafai2014connected} as well as in higher education, and it is also a sought-after skill in industry~\cite{hired}. 
Even though software testing is an integral part of software development in practice, it is often neglected in programming education~\cite{seth2014organizational}, despite increasing awareness in higher education and even proposals to include testing in the curricula of schools~\cite{DBLP:conf/acse/Carrington97, jones2001experiential, DBLP:conf/iticse/MarreroS05}. This is often caused by a lack of awareness and skills for teaching testing, and the perception shared by both learners and programmers that testing is a tedious and boring task. As long as software testing is not taught in a more engaging and accessible way, this is unlikely to change.

\begin{figure}
		\centering
		\includegraphics[width=\linewidth]{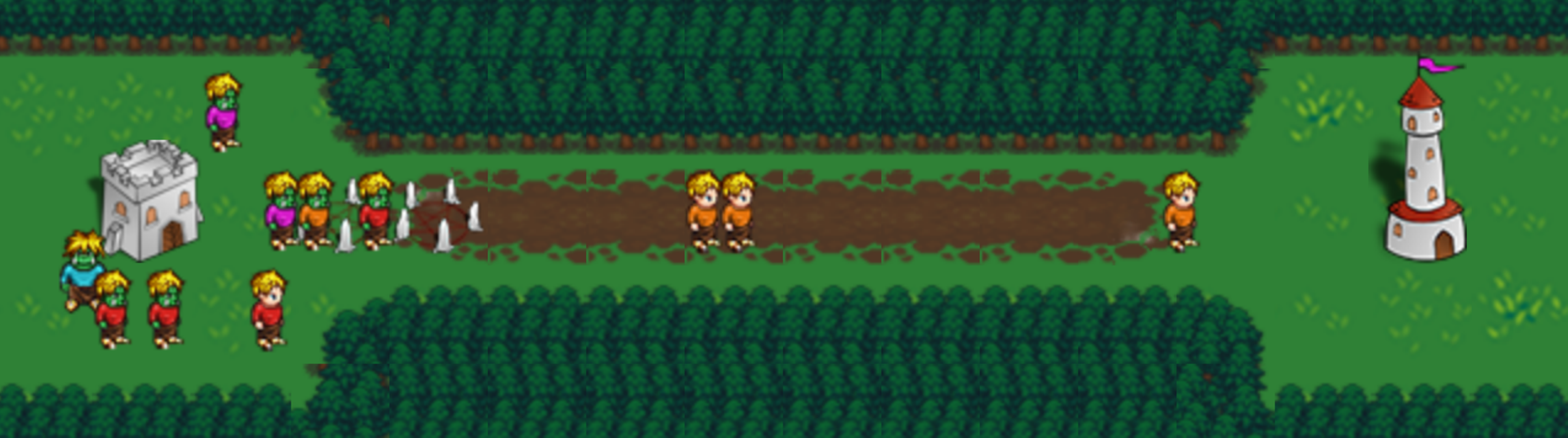}
		\caption{The gameboard during active gameplay}
		\label{fig:play}
	\end{figure}

To educate learners on testing in a more engaging way, we introduce
\toolname, a serious game based on Tower Defense, whose main purpose is not entertainment but education. \Cref{fig:play} shows the
gameboard in action: Players have to rescue human-like critters (for
example shown to the right in \cref{fig:play}) from mutants (critters
with zombie-like green heads further to the left), who can be
distinguished by their behavior encoded in small snippets of easily
accessible block-based code. The game is played by placing magic mines
(e.g., at the beginning of the dirt track in \cref{fig:play}), which
are essentially test cases, on the route the critters take from their
home on the left to the target tower on the right in
\cref{fig:play}. A mine represents test inputs (e.g., coordinates,
terrain type) as well as test oracles represented as block-based code
snippets. Different software concepts can be integrated into the
gameplay by building appropriate levels and mutants based on these
concepts.

\section{Background}

	Despite its importance and a growing awareness in higher education, software testing remains underrepresented during programming education~\cite{DBLP:journals/jss/GarousiRLA20}.
	A promising solution to incite developers to write tests is gamification (e.g.,~\cite{DBLP:conf/icse/StraubingerF22}), i.e., the use of game elements such as leaderboards, points or challenges in non-game contexts~\cite{DBLP:conf/mindtrek/DeterdingDKN11}. 
	Serious games take this approach a step further, as these are games explicitly made for training, education, or simulation, where players learn from embedded information about a topic without getting the feeling of learning or working \cite{DBLP:conf/chi/RaybournB05}. Serious games are often adaptations of well-known game types like Tower Defense, while changing facets to meet the objectives. Surprisingly, only few serious games have been proposed for software testing~\cite{DBLP:conf/icer/MiljanovicB17, DBLP:conf/icse/PrasetyaLMTBEKM19, DBLP:conf/fie/ToledoLS22}.

	One testing concept that has emerged as particularly suitable for gamification and serious games is mutation testing, which has, for example, been successfully integrated into the Code Defenders game~\cite{DBLP:conf/sigcse/FraserGKR19}. Mutation testing consists of inserting artificial defects in tested code to identify and remedy weaknesses in existing tests~\cite{5487526}. In Code Defenders, this is gamified as attackers creating artificial defects (mutants), which defenders aim to detect by writing tests. However, a central disadvantage of Code Defenders and other attempts to gamify testing is that they require reasonably advanced programming skills, and are therefore better suited for higher education. To also engage more inexperienced learners, testing must be integrated much earlier into programming education.

    A common approach to make programming accessible for younger learners is to represent code using block-based languages such as Scratch~\cite{maloney2010scratch}.
	Instead of typing code as text, learners assemble predefined code blocks visually by dragging and dropping them into position, quickly creating fun games and programs. Considering the success of block-based programming \cite{DBLP:journals/cacm/BauGKST17}, in this paper we explore the idea of similarly lowering the entry barrier also for software testing using a block-based programming approach.

\section{Code Critters}

	\toolname tells the story of the critters, who are people living in peace in an unknown colony in a forest. Unfortunately, a disease outbreak causes many of the critters to turn into mutants. These mutants do not behave like the other critters and are destroying the colony, forcing the healthy critters to flee into a tower across the forest. Along their walk to the tower, the player has to place magic mines, which check the behavior of the critters and only let healthy ones pass.
	
	\subsection{Game Concept}
	
	\begin{figure}
		\centering
		\includegraphics[width=0.8\linewidth]{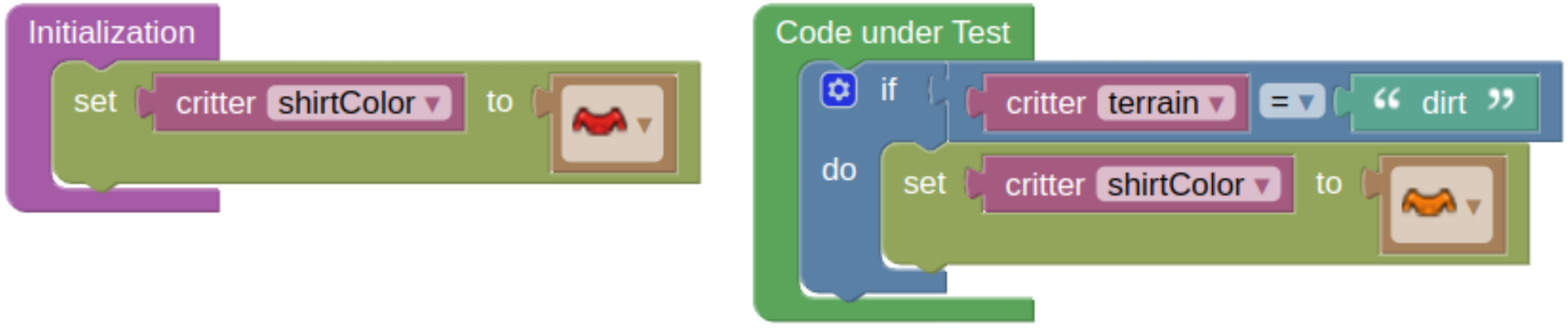}
		\caption{The critter under test}
		\label{fig:cut}
	\end{figure}
	
	\begin{figure}
		\centering
		\includegraphics[width=0.8\linewidth]{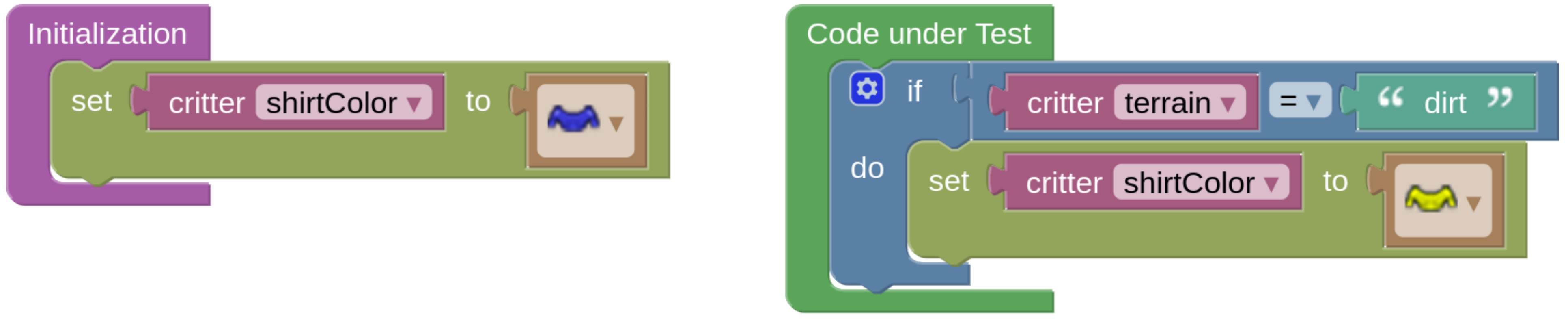}
		\caption{A mutant of the critter under test}
		\label{fig:mutant}
	\end{figure}
	
	\looseness=-1
	\toolname integrates mutation testing and block-based programming around the well-known game Tower Defense. In a classic Tower Defense game, the player has to place turrets along a route that enemies take to reach a certain point on the gameboard. The more enemies are eliminated on their way to this point, the more points the player receives in the end. Unlike traditional Tower Defense games, in \toolname there are not only enemies but also civilians who have to be protected. The behavior of critters is represented by the critter under test (CUT) as a short block-based code snippet (\cref{fig:cut}). Enemies (\emph{mutants}) are mutations of the CUT (\cref{fig:mutant}). To avoid the rather violent notion of killing enemies, in \toolname the turrets are replaced with mines that represent test cases; these mines use magic to trap the mutants instead of shooting them.
	
	\begin{figure}
		\centering
		\includegraphics[width=0.8\linewidth]{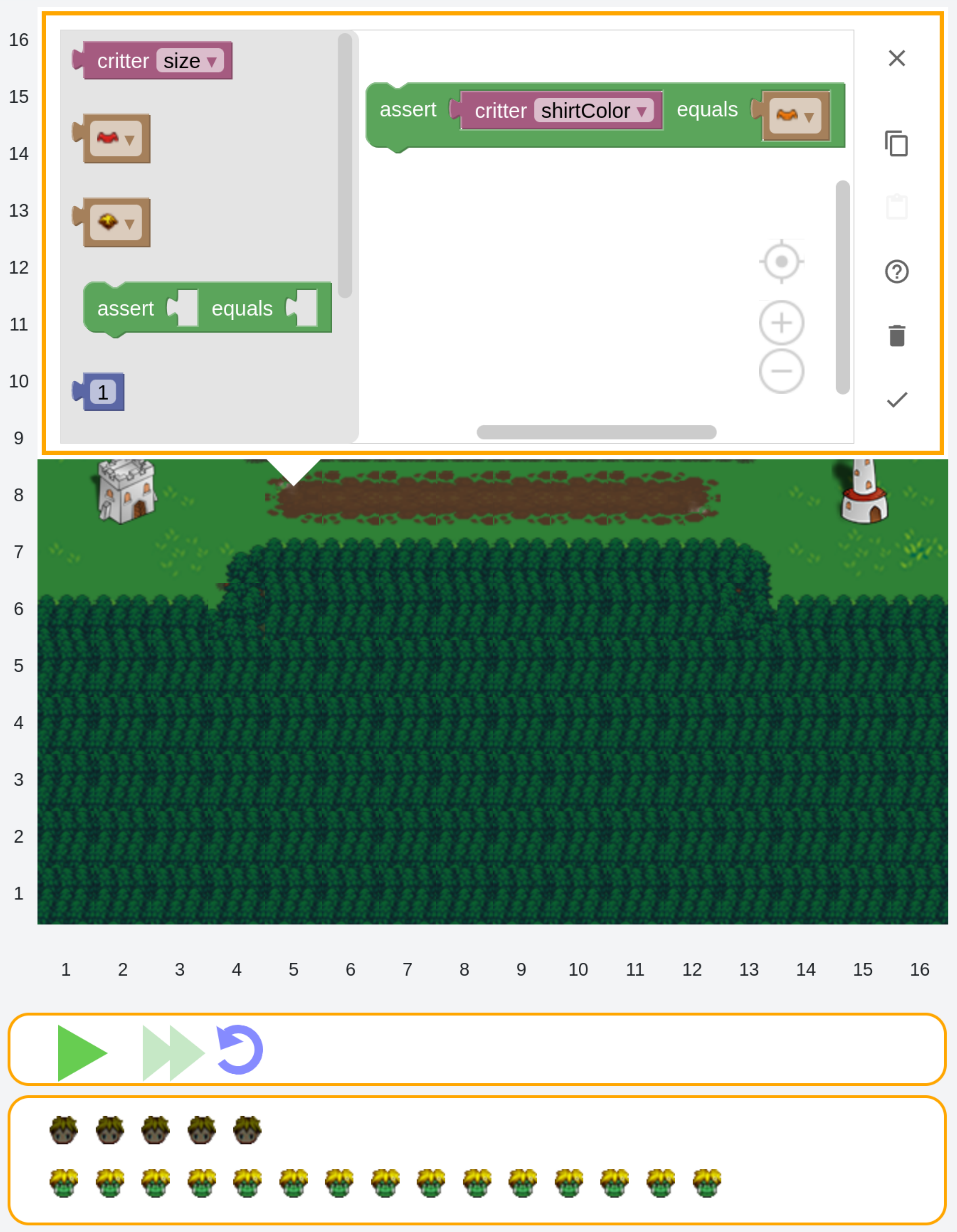}
		\caption{The gameboard of \toolname}
		\label{fig:game}
	\end{figure}
	
	\subsection{Game Mechanics}
	
	The gameboard (\cref{fig:game}) represents both the colony (the spawn point) and the tower (the destination), including the dirty trail from one to the other with the surrounding forest. The board is made of 256 tiles, represented with x and y coordinates from 1 to 16. Each tile can hold one specific texture, namely grass, dirt, water, ice, or wood, and critters can only walk on grass, dirt, and ice. In addition, mines can only be placed on tiles where critters can walk, which is also the first move the player has to make before starting the game.
	
	The behavior of critters is described with short snippets of code which are continuously executed in a loop while the critters are exploring the gameboard.
	A CUT is conceptually similar to an object oriented class and consists of two parts, the initialization and the code under test. The initialization behaves like a constructor and defines the initial values of the attributes of a critter like their shirt or hair color. The code under test is like a method that is called continuously in a loop while the critter walks, and receives texture and coordinates as inputs.
        A simple example of a healthy critter is the CUT shown in \cref{fig:cut}, which is initialized with a red shirt, and its color changes when walking on dirt tiles.

        Mutant critters contain one or more code mutations, such as the incorrect shirt color initialization and wrong choice of shirt color shown in \cref{fig:mutant}. The goal of the game is to let only healthy critters reach and enter the tower, while mutants are held off and trapped by the mines.

	Mines represent test cases: A mine is placed on exactly one tile by clicking on one of the walkable tiles, which represent the test inputs with their coordinates and texture.
	Clicking on a tile opens a dialog (\cref{fig:game}) in which a new test for the CUT can be created for a given input location (tile). Players can implement assertions for the mines with the same block-based language that represents the CUT or mutants. 
	The available blocks on the left side of the dialog in \cref{fig:game} can be placed on the right side via drag and drop and combined into a test case. The main block of each test is the assert-block, which is divided into the property to be checked and the value it should have at this point. Properties include the color of a critter's shirt and hair as well as their size. Shirt and hair colors are enumerations, while size properties are integers that can be checked for their value, whether they are even, odd, negative, positive or prime value. It is also possible to store and check basic information in variables and to perform basic mathematical operations on numerical values. The test represented by a mine is executed during active gameplay against any critter who steps onto the mine.
	
	To trap the mutant in \cref{fig:mutant}, two different mines have to be placed along the route to detect it: The first one somewhere on a grass tile which covers the mutation in the initialization, and the second mine on a dirt tile to find the mutation in the if-condition (see \cref{fig:play}). Both mines need to assert the expected value of the shirt color.
	These mines in combination with the mutants are also an abstraction layer to generate a test suite. A secondary goal of the game is to use as few mines as possible to detect all mutants, in other words, to minimize the test suite. In general, there is no limitation on the number of mines, but points are deducted if too many of them are used.
	
	\begin{figure}[t]
		\centering
		\includegraphics[width=0.4\linewidth]{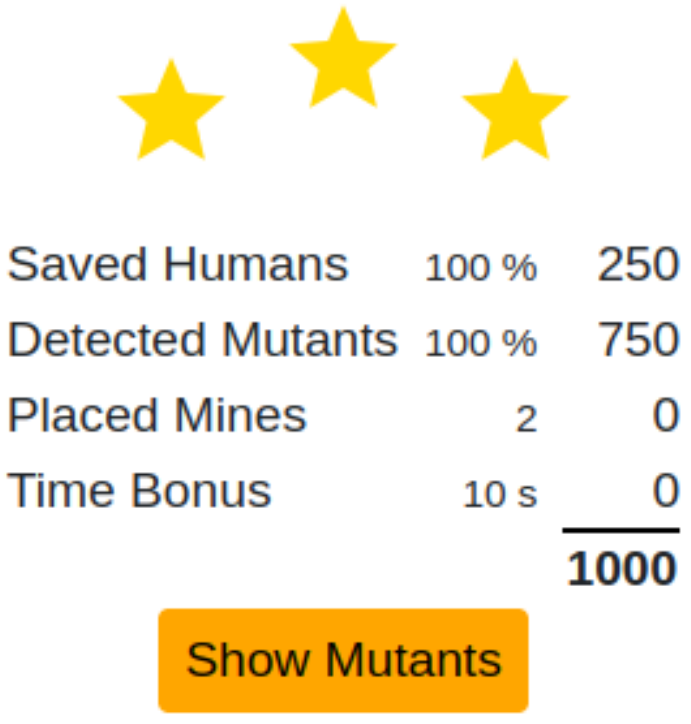}
		\caption{The scoreboard after finishing the current level}
		\label{fig:victory}
	\end{figure}
	
	After the player has placed all necessary mines on the board, the game can be started (\cref{fig:play}). The critters now start from the colony on the left and take a random route to the tower. A running game can be paused at any time as well as reset or sped up. A valid mine lets healthy critters pass without harm, while it traps mutants if the test fails (see \cref{sec:levels}). Consequently, invalid assertions may lead to false positives and trap healthy critters rather than mutants, which leads to the player losing points. At the bottom of the screen (\cref{fig:game}), the critters and mutants are displayed, and those who reached the tower will be marked as saved while trapped ones will be greyed out. When the last free critter or mutant reaches the tower, the game ends and the scoreboard is shown (\cref{fig:victory}). It contains information about saved critters and detected mutants, as well as the number of mines used and the given time bonus. After finishing the game, the mutants can be viewed to gain insights into how the CUT was mutated. The achieved score is accumulated with points earned from prior games and displayed on a public leaderboard on the starting page of \mbox{\toolname (\cref{fig:leveloverview})}.
	
	\subsection{Levels} \label{sec:levels}
	
	\begin{figure}
		\centering
		\includegraphics[width=0.7\linewidth]{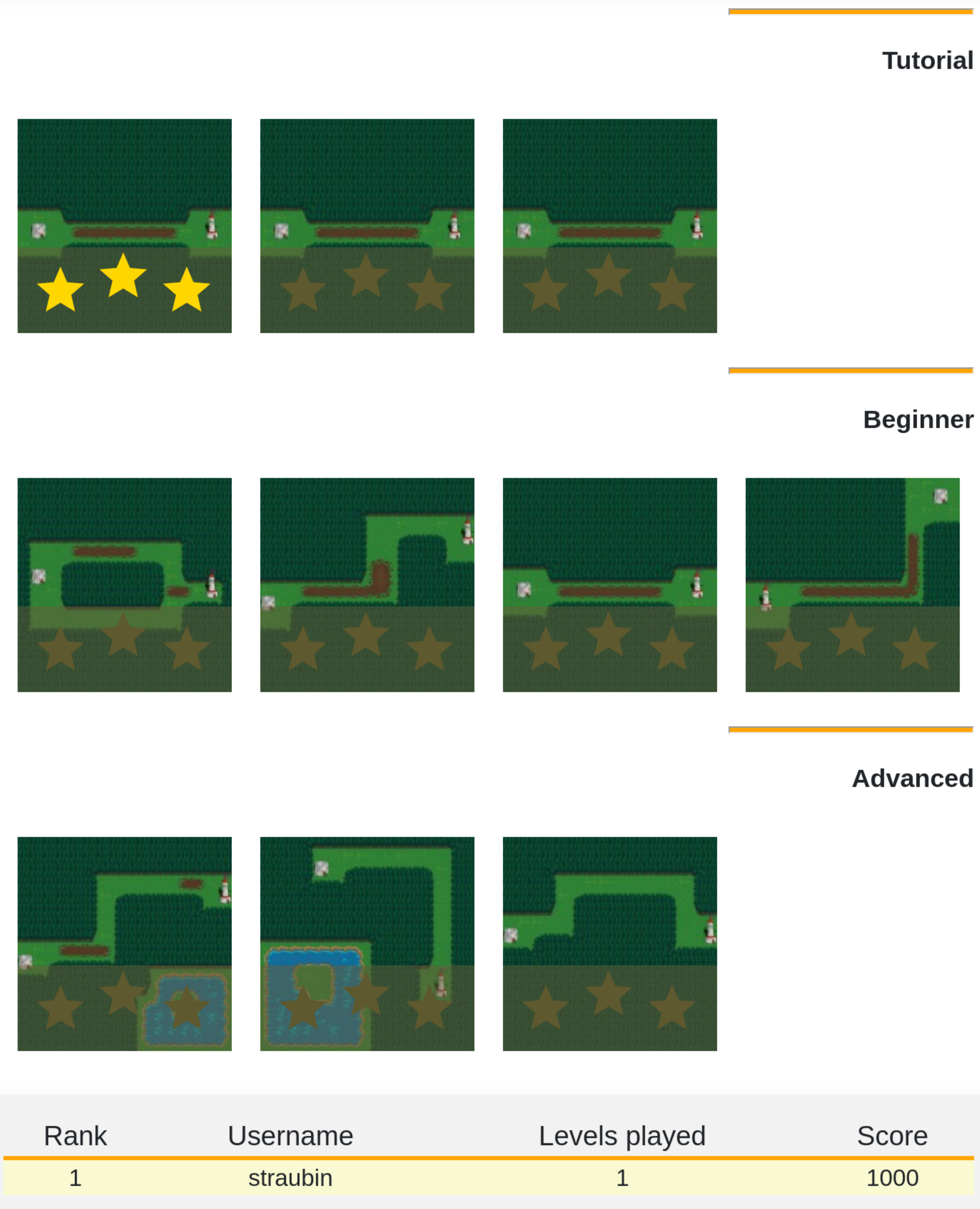}
		\caption{Different levels in \toolname with the leaderboard}
		\label{fig:leveloverview}
	\end{figure}
	
	\begin{figure}
		\centering
		\includegraphics[width=0.8\linewidth]{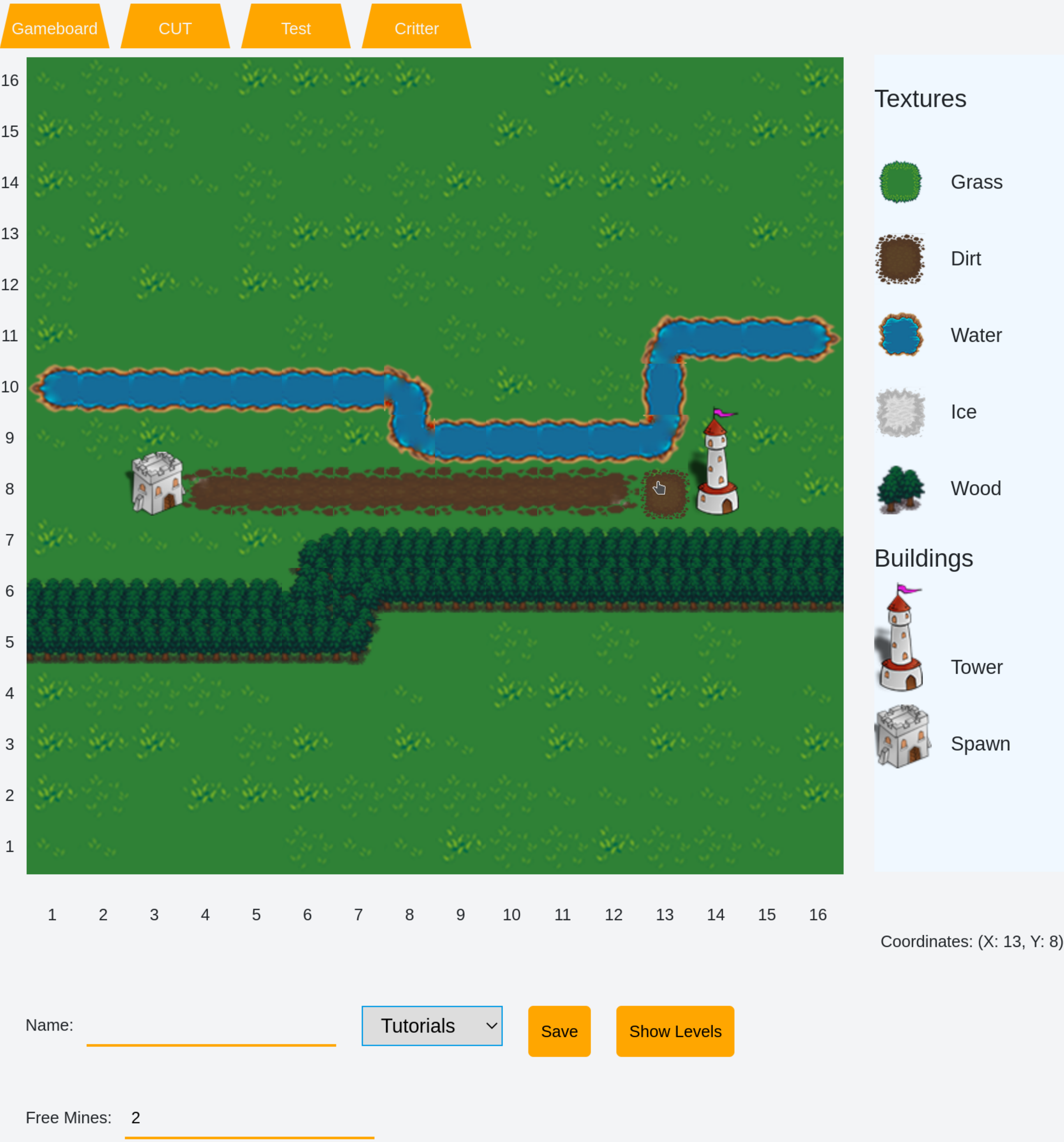}
		\caption{The level editor of \toolname}
		\label{fig:levelgenerator}
	\end{figure}

	\footnotetext[1]{\url{https://developers.google.com/blockly}}
	
	\toolname provides different levels, organized into tutorial, beginner, and advanced (\cref{fig:leveloverview}) categories. Our current proof-of-concept integrates ten levels, but \toolname also provides a level editor (\cref{fig:levelgenerator}) to create new ones. The level editor allows the creation of a custom map with the five different texture types and the start and end points. Each level needs a CUT to be created and, based on the CUT, one or more mutants. In addition, the number of critters, the number of mines the player can use without point deduction as well as the difficulty grade have to be set.
	
	Since each level is based on one CUT, the difficulty and learning goals can be adjusted by choosing an appropriate CUT.
        Players need to consider the following parts of the CUT and write tests for them to trap all mutants:
	
	\begin{itemize}
		\item \textbf{Initialization}: Mutation in the initial configuration of the critter
		\item \textbf{Assignments}: Changes in all property or variable assignments
		\item \textbf{Branches}: Removal or changes in if-else-clauses
		\item \textbf{Conditions}: Removal or changes of conditions in if-else-clauses
	\end{itemize}
	
	Levels can capture different testing concepts by adjusting those parts. For example, using a set of mutants where each mutant has one or more changed assignment statements, the concept of full statement coverage can be taught while changing the order or the content of if-else-clauses teaches about branch coverage. Changing or even removing conditions within conjunctions or disjunctions leads to the understanding of condition coverage.
	
	Not only the CUT can be mutated in various ways to increase the difficulty, but the gameboard itself can be altered and designed in a way to add more difficulty. For example, the board can be adjusted to provide more than one path the critters and mutants can take to the tower. This leads to the requirement for different mines on different tracks because of the texture and the coordinates of the tiles; some mines may be unique in one of the routes, and others have to be added in every route to ensure all mutants are caught, essentially teaching about input partitioning~\cite{collard2002practical}.
	
	\subsection{Implementation}
	
	\toolname is designed as a web application that can be deployed on any server and reached from the internet to be playable for everyone in a web browser. Like Scratch and many other block-based programming environments, we decided to build \toolname using the Blockly\footnotemark[1] library for representing and editing code.
	Scratch adds an abstraction layer over basic programming aspects by defining different blocks that the learners can combine into a meaningful program. In \toolname we reuse these concepts by defining the CUT with different blocks instead of source code (\cref{fig:cut}).

\section{Conclusions}

	\toolname is a proof-of-concept implementation that demonstrates the possibility to teach testing concepts with block-based programming in a fun way. \toolname is work in progress, and we plan to extend it with additional game elements, programming concepts, and corresponding levels in the future. It will also be important to study and evaluate \toolname with actual learners. The source code is available at:
	\url{https://github.com/se2p/code-critters}
        and \toolname can be tried out online at:
	\url{https://code-critters.org}

\section*{Acknowledgements}
This work is supported by the DFG under grant \mbox{FR 2955/2-1}, ``QuestWare: Gamifying the Quest for Software Tests''.

\balance
\bibliographystyle{IEEEtran}
\bibliography{IEEEabrv,bib}

\end{document}